\documentclass[superscriptaddress, nofootinbib, preprint]{revtex4}[12pt]
\usepackage{graphicx}
\graphicspath{{figs/}}
\usepackage {amsmath, amssymb, rotating}
\usepackage{hyperref}
\usepackage[usenames,dvipsnames]{color}
\usepackage{mdwlist} 
\usepackage{slashed}
\usepackage{verbatim}
\usepackage{mathrsfs}

\newcommand{\lmin}{\lambda_{\text{min}}}

\begin{document}

\title{Dark Matter Triggers of Supernovae}
\author{Peter W. Graham}
\affiliation{Stanford Institute for Theoretical Physics, Department of Physics, Stanford University, Stanford, CA 94305}

\author{Surjeet Rajendran}
\affiliation{Berkeley Center for Theoretical Physics, Department of Physics, University of California, Berkeley, CA 94720}

\author{Jaime Varela}
\affiliation{Berkeley Center for Theoretical Physics, Department of Physics, University of California, Berkeley, CA 94720}

\begin{abstract}
The transit of primordial black holes through a white dwarf causes localized heating around the trajectory of the black hole through dynamical friction. For sufficiently massive black holes, this heat can initiate runaway thermonuclear fusion causing the white dwarf to explode as a supernova.  The shape of the observed distribution of white dwarfs with masses up to $1.25 M_{\odot}$ rules out primordial black holes with masses $\sim 10^{19}$ gm - $10^{20}$ gm as a dominant constituent of the local dark matter density. Black holes with masses as large as $10^{24}$ gm will be excluded if recent observations by the NuStar collaboration of a population of white dwarfs near the galactic center are confirmed. Black holes in the mass range $10^{20}$ gm - $10^{22}$ gm are also constrained by the observed supernova rate, though these bounds are subject to astrophysical uncertainties. These bounds can be further strengthened through measurements of white dwarf binaries in gravitational wave observatories. The mechanism proposed in this paper can constrain a variety of other dark matter scenarios such as Q balls, annihilation/collision of large composite states of dark matter and models of dark matter where the accretion of dark matter leads to the formation of compact cores within the star. White dwarfs, with their astronomical lifetimes and sizes, can thus act as  large space-time volume detectors enabling a unique probe of the properties of dark matter, especially of dark matter candidates that have low number density. This mechanism also raises the intriguing possibility that a class of supernova may be triggered through rare events induced by dark matter rather than the conventional mechanism of accreting white dwarfs that explode upon reaching the Chandrasekhar mass. 
\end{abstract}

\maketitle

\section{Introduction}
\label{sec:intro}

Dark matter is proof of the existence of physics beyond the standard model. It is thus of great interest to identify its properties. The discovery reach of most dark matter experiments is limited to  candidates such as Weakly Interacting Massive Particles (WIMPs) and axions that are lighter than the weak scale. This limit on the reach exists because these experiments are typically laboratory scale devices  that operate over time scales of a few years. A large enough flux of dark matter needs to flow through this space-time volume in order for rare dark matter events to occur, limiting the range of masses that can be probed by such experiments. But, in light of our limited understanding of the physics of dark matter, there is no proof that the dark matter must be distributed around the universe as weak scale (or lighter) particles. While there are stringent bounds on interactions between the dark matter and the standard model (for example, see \cite{Aprile:2012nq}), interactions within the dark sector are not significantly constrained \cite{Wandelt:2000ad}. Much like self-interactions within the standard model producing composite objects whose masses range from that of a proton to supermassive black holes, self interactions within the dark sector can also produce complex composite states, with masses orders of magnitude larger than the weak scale \cite{Hardy:2014mqa}. Such dark matter states will have number densities orders of magnitude lower than that of conventional WIMP or axion dark matter, and can thus only be probed by detectors with a large space-time volume. 

We point out that white dwarfs, the progenitors of type 1a supernovae, can constrain scenarios that permit ultra-heavy dark matter states. These  include models where the dark matter accumulates in the white dwarf, with self-interactions allowing it to form a compact core inside the star and primordial black holes.  We show that if these objects are sufficiently massive, the heating induced by stellar matter attracted gravitationally to these dark matter objects is enough to cause the white dwarf to explode as a supernova. While this heating is significant only over a small region of the star, it triggers runaway fusion leading to an explosion. In this paper, we display this mechanism in detail for primordial black hole dark matter, leaving the study of dark matter models that can form a compact core inside the star for future work. As we will show, when a sufficiently massive primordial black hole goes through a  white dwarf, it will cause the white dwarf to explode even if the mass of the white dwarf is well below the Chandrasekhar limit. The observation of white dwarfs with masses below the Chandrasekhar limit can therefore be used to constrain the parameter space of such black holes. 

This mechanism allows  white dwarfs to act as dark matter detectors, with a collection area set by the size of a star $\sim 10^4$ km which is further increased by a gravitational Sommerfeld enhancement. This detector has been functional for time scales close to the age of the universe and when it is triggered, the subsequent supernova explosion is visible over astronomical distances, including observers on the Earth. This dramatically enhanced space-time volume enables this system to be sensitive to dark matter candidates such as primordial black holes that have a very low number density. 

We begin in section \ref{sec:overview} by presenting an overview of the explosion mechanism. Following this discussion, we compute the properties of black holes necessary to trigger this explosion in section \ref{sec:primordialBH} and use these properties to place constraints on the parameter space in section \ref{sec:observations}. We present our conclusions and ideas for future work in section \ref{sec:conclusions}. 


\section{Triggering Runaway Fusion}
\label{sec:overview}

The stellar medium of a white dwarf has conditions ripe to support runaway nuclear fusion. The white dwarf is  composed of materials such as carbon which can fuse through  strong interactions to form heavier elements. This fusion reaction requires the carbon nuclei to tunnel through a Coulomb barrier and is hence exponentially sensitive to temperature. An increase in the temperature of the nuclei by even factors of a few can dramatically enhance the rate of fusion \cite{Gasques:2005ar}. Every fusion reaction produces energy $\sim 10$ MeV - an energy high enough to efficiently overcome Coulomb barriers. Hence, the products of every fusion reaction have the ability to trigger even more fusion, potentially leading to a runaway explosion. Such a runaway reaction cannot occur in a medium like the Sun  (modeled roughly as an ideal gas), where a temperature increase causes a pressure increase that leads to expansion. This expansion decreases the density of the nuclear fuel and the fusion rate, thus achieving a stable burn rate. However, a white dwarf is supported by degeneracy pressure - wherein an increase in temperature does not affect the pressure. Hence, the stellar medium retains its density even as its temperature increases, preventing a regulation of the rate of fusion. The exponential increase of the nuclear fusion rate with temperature and the inability of the stellar medium to regulate its rate of fusion together makes the white dwarf capable of sustaining fusion chain reactions, leading to a thermonuclear runaway. Indeed, this is the underlying reason why white dwarfs explode as Type 1a supernovae when they come close to the Chandrasekhar mass. 

	This thermonuclear runaway can be triggered by sufficiently increasing the temperature of a small region in the star \cite{Timmes:1992}. The high temperature in that region increases the local rate of fusion. The energy released from these fusion reactions can initiate subsequent reactions leading to a chain reaction. But, when heat is introduced locally into the stellar medium, the heat will dissipate away from that region, lowering the temperature. This can dramatically decrease the fusion rate, causing the reaction to fizzle out. Hence, a chain reaction occurs as long as the rate of nuclear fusion is faster than the rate of thermal diffusion. 
	
	Thermal diffusion is governed by the heat equation. Consider a medium that is at an ambient temperature $T_0$. Now, increase the temperature of a region of size $\lambda$ by $T \gg T_0$.  The heat equation describes the evolution of this temperature profile. Using this equation, the time $\tau$ required for the temperature in the hot region to decrease by an $\mathcal{O}\left(1\right)$ fraction yields $\tau \approx \frac{c_{\rho} \rho}{K_{cd}} \lambda^2$ where $K_{cd}$, $\rho$ and $c_{\rho}$ are the conductivity, density and specific heat of the medium. The dissipation time scale gets longer as the initial size of the hot region becomes bigger. When this dissipation time scale is longer than the time required for nuclear fusion at temperature $T$, fusion will occur before the heat can dissipate. Since the fusion reaction is highly exothermic, the energy released from it will extend the size of the region that is at the high temperature $T$. The heat from this larger region will dissipate even more slowly while the fusion rate remains fixed enabling the nuclei in the larger region to fuse, resulting in a runaway \cite{Timmes:1992}. 

	The minimal size $\lmin$ that needs to be heated up to a temperature $T$ to trigger a thermonuclear runaway can thus be found by requiring the fusion rate $R(T)$ to be larger than the diffusion rate $\tau^{-1}$ \cite{Timmes:1992}. In addition to the expected exponential dependence on temperature, $\lmin$ is also a function of the stellar density. The fusion rate $R(T)$ is directly proportional to the density, while the thermal diffusion rate generally decreases with increased density \cite{Kippenhahn:1990}. Heat in the degenerate stellar medium can be efficiently transported by both photons and electrons, with the individual conductivities being strong functions of density and temperature \cite{Kippenhahn:1990}. For the range of parameters that is of most interest to us, both these conductivities are relevant, complicating analytical treatments of the problem. However, the problem can be tackled numerically and this was done in \cite{Timmes:1992}. 

	Before discussing these numerical results, let us first understand the analytic dependence of $\lmin$ on density. The conductivity $K_{cd} \propto \frac{T^3}{\kappa_{cd} \, \rho}$ where $\kappa_{cd}$ is the opacity of the carrier. At densities and temperatures relevant to us, the photon opacity is dominated by free-free absorption which scales with density as $\kappa_{cd} \propto \rho$  \cite{Kippenhahn:1990}. In degenerate matter, electron opacities are also relevant due to the absence of soft scattering and this opacity scales with density as $\kappa_{cd} \propto \rho^{-2}$. In agreement with the results of \cite{Timmes:1992}, we find that for densities larger than $\sim 10^{8} \frac{\text{gm}}{\text{cm}^3}$ and temperatures greater than 500 keV, electrons are the dominant heat carrier. Estimating $\lmin$ from the condition that $R(T) \tau \gtrapprox 1$, in this regime we find  $\lmin \propto \rho^{-\frac{1}{2}}$. At lower densities and temperatures $\gtrapprox$ 500 keV, photons become the dominant heat carrier, yielding $\lmin \propto \rho^{-2}$. In both cases, the trigger size $\lmin$ becomes larger as the density drops. We use these analytic approximations to scale the results of  \cite{Timmes:1992} over a wider range of densities than calculated by them. 

	For a range of initial densities, the authors of  \cite{Timmes:1992}  calculated the minimum size $\lmin$ of a spherical region that must be heated to a temperature $T$ in order to create a thermonuclear runaway \cite{StanPrivate}. For example, they find that for densities  $\sim 5 \times 10^9 \frac{\text{gm}}{\text{cm}^3}$, a region of size $\sim 10^{-5}$ cm needs to be heated to a temperature $\sim$ MeV to trigger the runaway. This size increases to $\sim 10^{-4}$ cm for stellar densities   $\sim 2 \times 10^8 \frac{\text{gm}}{\text{cm}^3}$. These trigger sizes are valid for carbon-oxygen white dwarfs and are about $\sim 10$ larger for oxygen-neon-magnesium dwarfs. These computations are consistent with the expected analytic behavior of $\lmin$. The authors of  \cite{Timmes:1992}  did not compute the trigger sizes for densities below $\sim 10^8 \frac{\text{gm}}{\text{cm}^3}$. In this regime, we scale their results using the  analytic scaling of $\lmin$. We note that the conclusions of \cite{Timmes:1992}   on the minimum trigger size is close to the naive order of magnitude estimate of the size obtained by setting the fusion rate equal to the diffusion rate $R(T) \tau \approxeq  1$ (the authors of  \cite{Timmes:1992}  have also noted this point), adding confidence to our analytic estimate of the trigger size.  To arrive at this estimate, we used nuclear reaction rates from \cite{Gasques:2005ar, Caughlan:1988} and thermal conductivities from \cite{Kippenhahn:1990}. 

	Additionally, as noted in  \cite{Timmes:1992}  (and easily verified by solving the heat equation), the diffusion rate does not change appreciably by changing the geometry of the initial hot region from being a sphere of size $\lmin$ to that of a cylinder of radius $\lmin$. Consequently, any effect that can increase the temperature of a region of size $\lmin$  to $\approx$ 1 MeV inside the white dwarf will cause it to explode. In the next section, we show that sufficiently heavy primordial black holes can accomplish this feat and are thus constrained. 


\section{Primordial Black Holes}
\label{sec:primordialBH}
\begin{figure}[h]
\centering
 \includegraphics[width=5in]{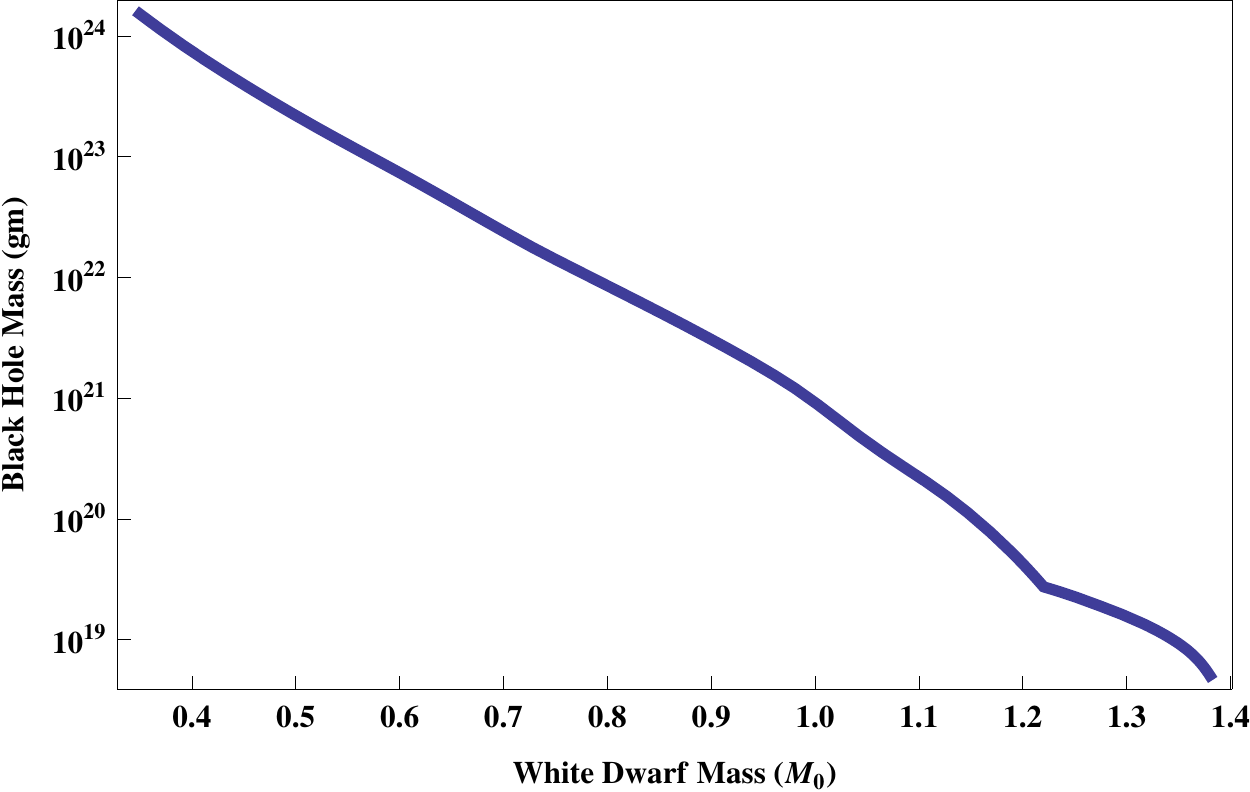}
\caption{The minimum mass of a black hole whose transit can destroy a carbon white dwarf of a given mass. The black holes transit the star with a speed equal to the escape velocity of the star, estimated using the mass-radius relationship of the white dwarf to be $\approx 2 \times 10^{-2}$. The plot uses a digitization of the equation of state of the white dwarf from \cite{TimmesCode} to relate the density of the white dwarf to its mass. The density is nearly constant for much of the star and hence off-center transits of black holes are also equally explosive.  \label{figdestroyer}}
 
\end{figure}
Primordial black holes are intriguing dark matter candidates. While it is challenging to produce them through conventional standard model processes (see \cite{Belotsky:2014kca, Linde:2012bt} for one mechanism), it is possible that new physics in the sector responsible for dark matter may produce them. There are  constraints on such black holes being dark matter - if they are lighter than $\lessapprox 10^{17}$ gm,  Hawking emission from them contributes too much to the observed cosmic ray spectrum. Black holes heavier than $\sim 10^{24}$ gm are constrained by lensing data \cite{Pani:2014rca} and Kepler observations \cite{Griest:2013aaa}. Attempts have been made to place bounds between  $\sim 10^{17}$ gm - $10^{24}$ gm \cite{Pani:2014rca, Barnacka:2013ana, Capela:2014ita}, but are subject to  uncertainties \cite{Pani:2014rca, Defillon:2014wla}.

	Suppose a black hole with a mass in the presently allowed region  ($\sim 10^{17}$ gm - $10^{24}$ gm) goes through a white dwarf. These black holes are small enough (Schwarzschild radii  between $\sim 10^{-9}$ cm - $10^{-4}$ cm) that they will simply go through the star.  The purely gravitational effects of their passage through the star is negligible and they do not lose enough momentum through collisions with stellar matter to stop (or significantly slow down) in the star. However, as they pass through the star, they will accelerate stellar matter heating a cylindrical region around their trajectory. If this transit succeeds in heating a region of size $\gtrapprox \lmin$ by $\sim$ 1 MeV, it will trigger thermonuclear runaway, leading to an explosion. 

	The heat induced by the transit can be calculated from dynamical friction \cite{Capela:2013yf}. The primordial black hole goes through the star with a velocity $v_{\chi}$ marginally greater than the escape velocity, which is also $\sim 10^{-2}$ for a white dwarf. This velocity is thus greater than the sound speed in the star and hence the perturbation caused by the black hole's transit can be calculated using the sudden approximation. In this limit, the black hole imparts an instantaneous velocity to the carbon nuclei around it. To calculate this velocity, note that a black hole of mass $M_{\chi}$ moving through the star  accelerates a carbon nucleus at a distance $R$ from it by an amount $\sim \frac{G M_{\chi}}{R^2}$. This acceleration is effective for a transit time $\sim \frac{R}{v_{\chi}}$. Hence, the velocity induced by the transit is $\sim \frac{G M_{\chi}}{R v_{\chi}}$. These velocities are directed towards the trajectory of the black hole, leading to a situation where nuclei on opposite sides of the trajectory are aimed towards each other with this velocity. The ensuing collisions will result in the thermalization of the energy imparted to the nuclei by the black hole. To heat carbon nuclei to a temperature $\sim$ MeV, this velocity change should be $\sim 10^{-2}$. Equating these, we see that the black hole heats a cylindrical region of radius $R \sim 10^4 \, G M_{\chi} \sqrt{\log\left(R/GM_{\chi}\right)}$ to a temperature $\sim$ MeV. The logarithmic enhancement is $\approx 10$, yielding a modest increase to the radius $R$.  From this estimate, we see that black holes with a mass $M_{\chi} \gtrapprox 10^{19}$ gm will heat up cylindrical regions with radius larger than $\sim 10^{-5}$ cm to temperatures $\approx$ MeV. As per the discussion in section \ref{sec:overview}, the transit of such black holes will trigger supernova explosions in white dwarfs. 

	The trigger size to initiate runaway fusion depends upon the density of the white dwarf. The density of the star is determined by its mass through its equation of state. We use the results of \cite{TimmesCode} to estimate this relationship. With this functional dependence, for a white dwarf of a given mass, we calculate the minimum mass that a black hole must have in order for its transit to destroy the star. These results are plotted in figure \ref{figdestroyer}.





\section{Constraints}
\label{sec:observations}
\begin{figure}[h!]
\centering
 \includegraphics[width=5in]{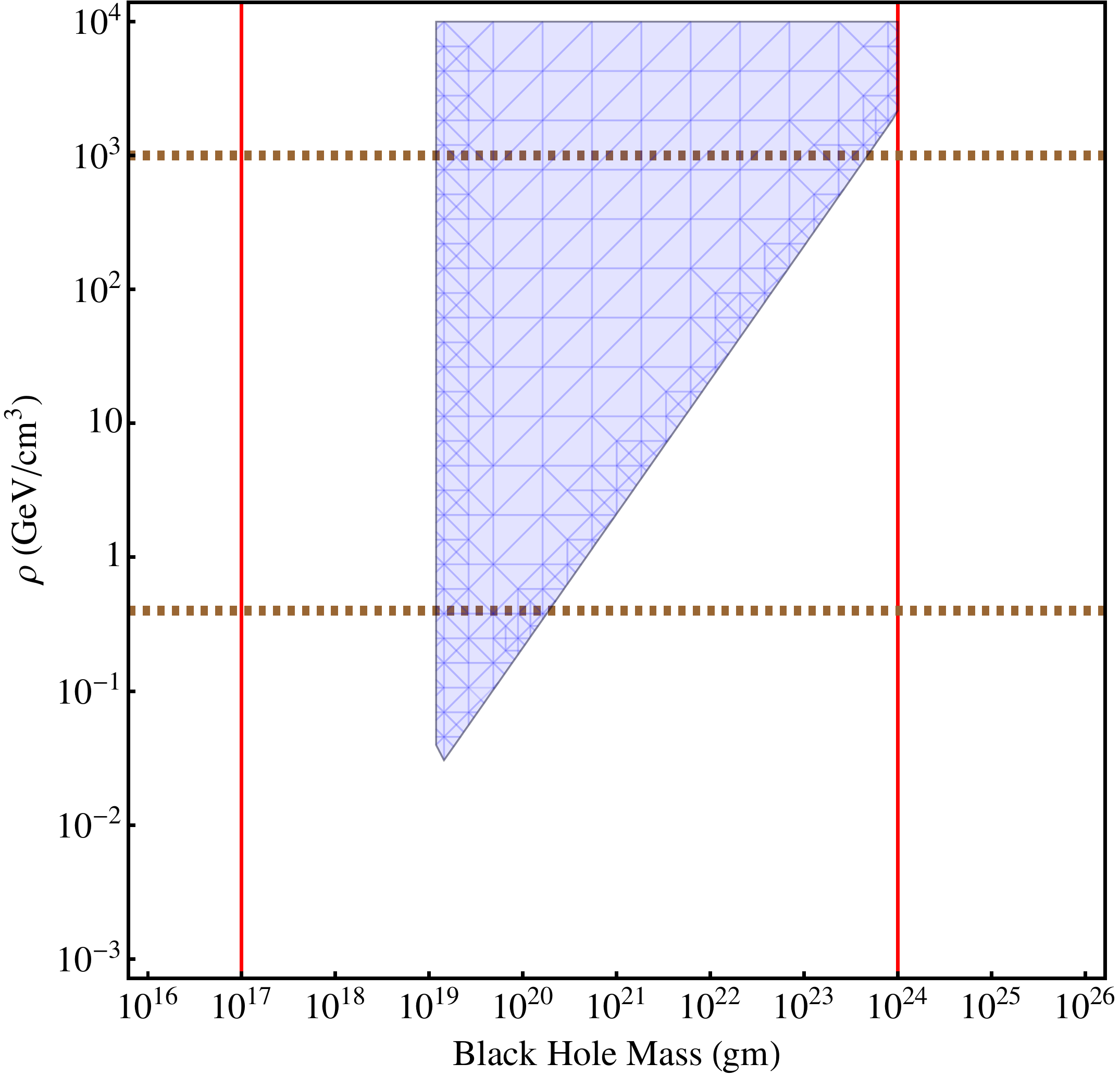}
\caption{The minimum dark matter density of primordial black holes as a function of its mass for one black hole to transit a white dwarf in 1 Gyr, with the minimum black hole mass of interest set by the mass necessary to destroy  RX J0648.04418. To estimate the gravitational capture cross-section, the white dwarf radius was taken to be $\approx 3000$ km with a stellar escape velocity of $\approx 2 \times 10^{-2}$. These were obtained using the mass-radius relationship of the star. While these numbers vary as a function of the mass of the star, the variation is not significant. The black holes were assumed to have virial velocities $\approx 10^{-3}$. \label{figdensities}}
 
\end{figure}

\begin{figure}[h]
\centering
 \includegraphics[width=5in]{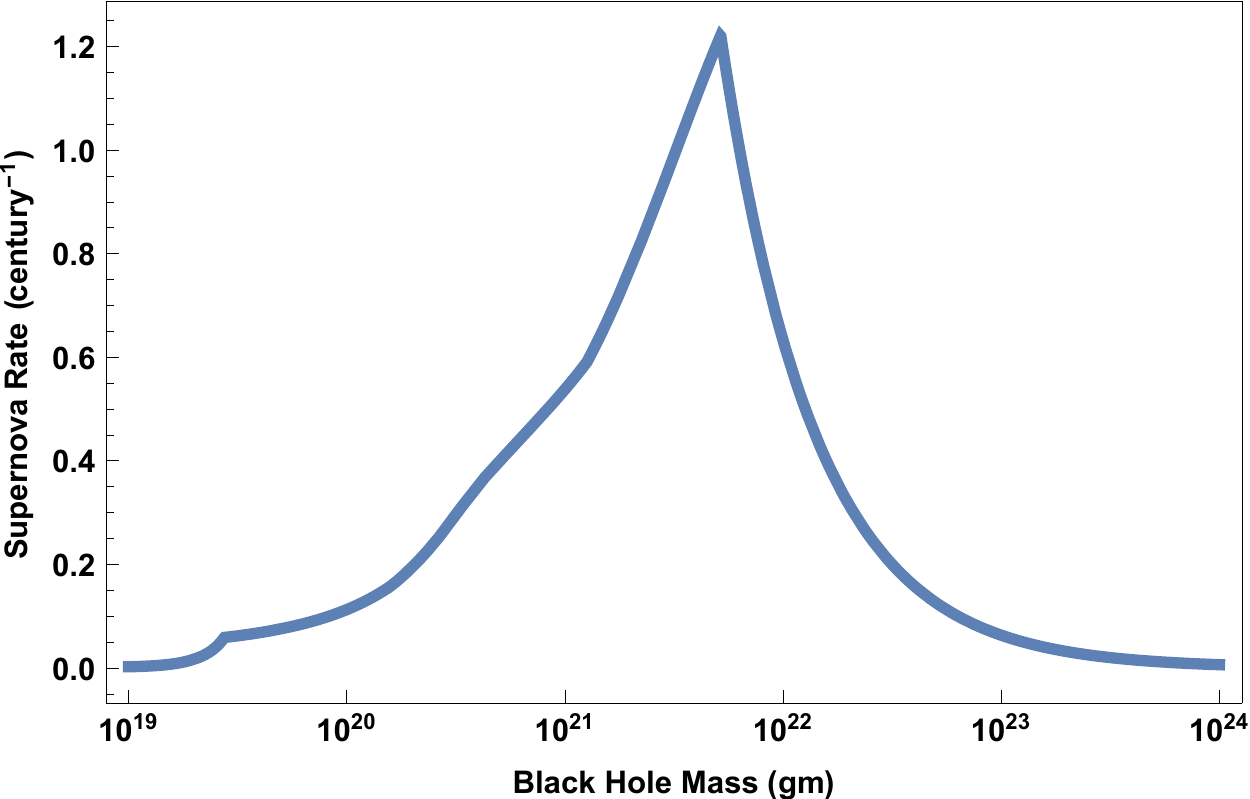}
\caption{The supernova rate as a function of primordial black hole mass under the assumption that the black hole density is equal to the local dark matter density $\approx 0.4 \text{ GeV/cm}^3$. The assumptions used to calculate this rate and the attendant caveats are described in the text. \label{figsupernova}}
 
\end{figure}

We use two classes of observations to place bounds on primordial black holes. For these bounds to constrain primordial black hole parameter space, the black holes need to traverse white dwarfs that are sufficiently heavy (as estimated in figure \ref{figdestroyer}). One class of bounds arises from direct observations of white dwarfs whose masses have been measured. Their existence constrains primordial black holes that are abundant enough that they would have transited the white dwarf with high probability. A second class of bounds arises from the measured rate of type 1a supernova. These can constrain primordial black holes with a low abundance, wherein they have a low probability of transit through a given white dwarf but may nevertheless destroy enough of them to produce too many supernovae. 

	Both cases require the computation of the transit rate for primordial black holes through a white dwarf. The cross section for primordial black holes to transit the white dwarf is given by the gravitational capture cross section of the star which is equal to $\sim \pi R_s^2 \left(v_{\text{esc}}/v\right)^2$ where $v_{\text{esc}}$ is the escape velocity of the star and $v \approx 10^{-3}$ is the virial velocity of primordial black hole dark matter. The transit rate $\Gamma_{\chi}$ is equal to $\sim  \left(\rho_{\chi}/M_{\chi}\right) \, \pi R_s^2 \left(v_{\text{esc}}^2 /v\right)$.   While this rate increases as the black hole mass decreases, the minimum mass of the white dwarf that can be destroyed by the transit increases. While most white dwarfs have masses $\sim 0.6 \, M_{\odot}$, there are a few known systems such as RX J0648.0Ð4418 that have masses  $\gtrapprox 1.25 \, M_{\odot}$ \cite{Mereghetti:2013nba}. This star is in a binary system, accreting matter from its companion and is thus likely to be a carbon white dwarf.  This star can be destroyed by primordial black holes with masses $\gtrapprox 10^{19}$ gm (see figure \ref{figdestroyer}), setting the minimum primordial black hole mass that can be probed by our method. RX J0648.0Ð4418  is by no means the only such heavy white dwarf - for example, the Sloan Digital Sky Survey \cite{Sloan} has observed 17 other white dwarfs with  masses known to be larger than 1.25 $M_{\odot}$. 

The first class of bounds is set by direct observations of white dwarfs.  For a given black hole mass, all white dwarfs above a certain critical mass can be blown up. Let $\tau_\text{BH}$ be the expected time for such a  primordial black hole to encounter a white dwarf, calculated assuming a cosmic density of such black holes. Suppose $\tau_\text{BH} \ll t_0$, where $t_0$ is the age of the universe (and the expected age of the white dwarf). Observations of white dwarfs with masses above the critical mass can be consistent with the assumed cosmic abundance of such black holes only if there was a significant enhancement in the  population of white dwarfs above this critical mass. However, such an enhancement is in contradiction with the observed smoothly falling distribution of white dwarfs with masses just below the critical mass. To be conservative, we consider primordial black holes that encounter a white dwarf on a time scale $\tau_\text{BH} \sim 1$ Gyr. The observed population of white dwarfs with masses above the appropriate critical mass and estimated age $t_0 \gtrapprox 5$ Gyr constrains the cosmic abundance of such black holes. We plot these bounds in Figure \ref{figdensities}.

	The lower, horizontal dotted line in figure  \ref{figdensities} corresponds to the mean galactic dark matter density of $0.4 \text{ GeV/cm}^3$. Since most white dwarfs we have seen, such as RX J0648.0Ð4418,  are located close to the Sun, this is a good estimate of the dark matter density around them. From  figure \ref{figdestroyer}, their existence rules out primordial black holes in the mass range $10^{19} \;\text{gm - } 10^{20} \;\text{gm}$. While heavier black holes can destroy lighter white dwarfs,  these white dwarfs need to be observed in regions of high dark matter density such as the galactic center (corresponding to the upper, horizontal dotted line in figure  \ref{figdensities}). Observations of faint white dwarfs in the galactic center is difficult at optical frequencies. However, measurements can be made at X-ray frequencies when the white dwarf is accreting matter. The NuStar collaboration has recently uncovered evidence for the existence of such a population of white dwarfs in the galactic center with an average mass as large as $1.2 M_{\odot}$ \cite{BoggsPrivate}. Confirmation of these observations would rule out primordial black holes with masses between $10^{20}$ gm - $10^{24}$ gm as a significant component of dark matter. 

	The explosions of white dwarfs caused by the transit of primordial black holes will contribute to the rate of type 1a supernova. Heavier black holes, while fewer in number, have the potential to explode lighter white dwarfs that are more abundant in number. However, for the explosion to be optically visible it must produce sufficient quantities of Ni 56. Simulations suggest that at least $\sim 0.85 \, M_{\odot}$ of material must be burnt in order for the yield of Ni 56 to be large enough to make the explosion visible optically \cite{KasenPrivate, Kim:2010}. White dwarf distributions such as  \cite{Gianninas:2009} and \cite{Sloan} suggest that about 5\% of all white dwarfs have masses larger than $\sim 0.85 \, M_{\odot}$. Using these distributions to estimate the fraction of white dwarfs above a given mass and the   expected number $\sim 10^{10}$ white dwarfs per galaxy \cite{Napiwotzki:2009}, we estimate the contribution of these transits to the type 1a supernova rate and plot the results in figure \ref{figsupernova}. The peaked structure of figure  \ref{figsupernova} is reflective of the fact that while the less abundant, heavier black holes can destroy lighter and more abundant white dwarfs, this destruction contributes to the type 1a supernova rate only when the white dwarf mass is larger than $\sim 0.85 \, M_{\odot}$. 

	The results of figure \ref{figsupernova} should be compared to the observed rate of type 1a supernova of 0.3 - 0.4 per century\cite{Bergh:1991}.  The type 1a rate caused by primordial black holes with a dark matter density 0.4 GeV/$\text{cm}^3$ and a mass between $\sim 10^{20}$ gm - $10^{22}$ gm is comparable to the observed rate. Since the white dwarf population falls steeply with mass, most of the stars exploded by these black holes have masses between $\sim 0.85 \, M_{\odot} \, - \, 1.0 \, M_{\odot}$. These explosions are weaker than those caused by the explosion of Chandrasekhar mass white dwarfs (which explode at $\sim 1.38 \, M_{\odot}$) and will have lower Ni 56 yields. Hence, these transits cannot be the cause of the observed type 1a rate and are thus constrained. However, there are important caveats to this constraint. First, the calculation of the supernova rate relies upon the population of white dwarfs with masses larger than $\sim 0.85 \, M_{\odot}$. This population is estimated using the high mass tail of the white dwarf distribution in the solar neighborhood and estimates of the total number of white dwarfs in the galaxy. There is thus  uncertainty in this number. Further, the estimate that the white dwarf must be at least as heavy as $\sim 0.85 \, M_{\odot}$ to produce enough Ni 56 is based upon current simulation data \cite{Kim:2010}. With refinements, it is possible that this number can shift. Since this number sets the high mass tail of the white dwarf population that contributes to the type 1a rate, changes in it can significantly affect the estimated rate. In light of these uncertainties, while the type 1a rate disfavors black holes with masses between $\sim 10^{20}$ gm - $10^{22}$ gm, their exclusion is not robust.

\section{Conclusions}
\label{sec:conclusions}

White dwarfs can explode if a localized region of the star becomes sufficiently hot. Using this fact, we have shown that their existence can be used to constrain primordial black hole dark matter, whose transit through the star gravitationally heats a region around its trajectory. The shape of the observed distribution of white dwarf masses excludes primordial black holes with masses $\sim 10^{19}$ gm - $10^{20}$ gm that have an abundance equal to the local dark matter density. Heavier black holes can be constrained if white dwarfs are confirmed to exist in regions of high dark matter density such as the galactic center. If preliminary results from the NuStar collaboration of a population of white dwarfs with masses $\sim 1.2 M_{\odot}$ near the galactic center are confirmed, primordial black holes with masses $\sim 10^{20}$ gm - $10^{24}$ gm cannot be a significant component of dark matter. Black holes in the range $10^{20}$ gm - $10^{22}$ gm can also contribute significantly to the observed rate of type 1a supernova. Owing to uncertainties in the white dwarf population distribution, while this rate disfavors the existence of such black holes it does not present a robust exclusion. 

Robust detections of white dwarf binaries  in which one of the component stars has a mass larger than $\sim 1.2 M_{\odot}$ with the binary located in a region of high dark matter density such as the galactic center will exclude  primordial black hole dark matter with masses above $\sim 10^{20}$ gm. While  there are challenges in optically observing white dwarf binaries in regions such as the galactic center, gravitational wave observatories such as the AGIS  \cite{Graham:2012sy, Dimopoulos:2008sv} 
and LISA \cite{LISAPrePhaseA} proposals are expected to be sensitive to such binaries. Using a combination of electromagnetic and gravitational wave measurements, the parameters of the binary systems can be well determined \cite{WhiteDwarfGW}, enabling a robust exclusion. Further, these observatories are also expected to be sensitive to stochastic sources of gravitational waves from unresolved extragalactic white dwarf binaries. This can help establish the white dwarf binary population outside our local galactic region, placing our estimate of the supernova rate on firmer footing. 

	While the focus of this paper was restricted to primordial black holes, the mechanism proposed by us can be used to constrain any model of dark matter that can provide localized heating. This includes examples such as Q-ball dark matter whose transit through the star causes heating by standard model processes. Collisions and annihilations of large composite dark matter states occurring inside the white dwarf can also cause localized heating. Further, dark matter particles may accumulate in white dwarfs.  If the dark matter forms a compact core within the star, with the density of the core dominating over the local baryon density, there will be a small region around this core where the baryons will be gravitationally heated to temperatures larger than their ambient values. If this region is sufficiently big, it will cause the white dwarf to explode. Bounds on such possibilities have been placed when such a compact core forms inside a neutron star and  becomes sufficiently massive to form a black hole \cite{McDermott:2011jp}. Our probe does not require the formation of a black hole - in fact, the white dwarf will explode well before the compact core becomes a black hole since the explosion occurs once the baryons near the core are heated to $\sim$ MeV temperatures as opposed to the case of the black hole where the baryons will be at $\sim$ GeV temperatures. Being larger in size, white dwarfs will also accumulate more dark matter than neutron stars. Additionally, the ability of dark matter to form a compact core inside neutron stars is hindered by the expected superfluidity of the neutron star core that inhibits soft scattering necessary to form such a core \cite{Bertoni:2013bsa}. Such an inhibition is not expected in white dwarfs. For these reasons, the process described in this paper is expected to place new constraints on such scenarios. We study these matters in further detail in a forthcoming publication. 

	In addition to placing bounds, the mechanism suggested by us raises the intriguing possibility that a class of type 1a supernovae may be triggered through the accretion of dark matter rather than the conventional assumption that such supernovae are caused by white dwarfs close to the Chandrasekhar limit. If this was the case, there would be a correlation between the supernova rate and the dark matter density in different locations.  Since such supernovae will be caused by stars with masses well below the Chandrasekhar limit, they will have varying luminosities, providing an interesting handle to isolate such events, and possibly an explanation for some of the observations of supernova luminosities. If the mechanism responsible for the explosion is an off-center heating event such as the transit of a primordial black hole or a Q-ball, or the collision/annihilation of composite dark matter states, then the explosion will not have an axis of symmetry, potentially leading to the production of gravitational waves.   Further, this may change the prediction for the elemental abundances created by the supernova.  It would be interesting to calculate the predictions of this intriguing possibility to see how well it fits observations.

	The standard model exhibits an extraordinary range of complexity. There is no reason to think that the dark matter may also not be similarly complex. Such complexities invariably result in the production of ultra-massive states, whose vastly reduced number density hinders our ability to probe them with conventional human scale experiments. White dwarfs, with their astronomical lifetimes and sizes, can overcome these limitations and offer a unique ability to probe complex dark sectors. 


\section*{Acknowledgements}
We would like to thank D. Finkbeiner, S. Kachru,  J. Kalirai, D. Kasen, A. Kusenko,  K. Makishimia, R. Romani, K. Shen, D. Wang and S. Woosley
for many useful discussions.  PWG acknowledges the
support of NSF grant PHY-1316706, DOE Early Career Award
DE-SC0012012, and the Terman
Fellowship. SR acknowledges the support of NSF
grant PHY-1417295. JV acknowledges the support of NSF
grant DGE-1106400.

\end{document}